\documentclass[aps,pra,preprint]{revtex4-1}

\usepackage{amsmath,amssymb,graphicx,color}

\newcommand{\eqn}{\begin{equation}}
\newcommand{\nqe}{\end{equation}}
\newcommand{\bea}{\begin{eqnarray}}
\newcommand{\eea}{\end{eqnarray}}
\newcommand{\nn}{\nonumber}

\newcommand{\bra}[1]{\langle #1 \rvert}
\newcommand{\ket}[1]{\lvert #1 \rangle}

\newcommand{\abs}[1]{\lvert #1 \rvert}
\DeclareMathOperator{\rmd}{d}
\DeclareMathOperator{\sgn}{sgn}

\DeclareMathOperator{\erfc}{erfc}

\numberwithin{equation}{section}

\begin{document}

\title{Quench dynamics in integrable systems} \author{Natan Andrei}
\affiliation{Department of Physics\\ Rutgers University\\
  Piscataway, New Jersey 08854.}  \date{\today}

\begin{abstract}
These notes cover in some detail  lectures I gave at the Les Houches Summer School  2012. I describe here work done  with Deepak Iyer with important contributions from Hujie Guan. I discuss some aspects of the physics revealed by quantum quenches  and  present  a formalism for studying the quench dynamics of
  integrable systems. The formalism presented  generalizes an approach by Yudson and is  applied  to
   Lieb-Liniger model which  describes a gas of  $N$ interacting bosons
  moving on the continuous infinite line while  interacting via a short
  range potential.  We carry out the quench from several initial states and for any  number of particles and compute the evolution of the density and the noise correlations. In the long time limit the system dilutes and we find that for  any value of repulsive coupling
  independently of the initial state the system asymptotes towards a
  strongly repulsive gas, while for any value of attractive coupling,
  the system forms a maximal bound state that dominates at longer
  times.  In either case the system equilibrates but does not
  thermalize, an effect that is consistent with prethermalization. These results can be confronted with experiments. For much more detail see: Phys. Rev. A 87, 053628 (2013) on which these notes are based.
  Further applications of the approach to the Heisenberg model and to the Anderson model  will be presented elsewhere.
\end{abstract}

\maketitle

\section{Introduction}
\label{sec:intro}

Over the past recent years the study of nonequilibrium dynamics of isolated quantum systems has
seen significant advances both in theory and experiment. Although nonequilibrium processes abound in all fields of science and in particular in condensed matter physics, their detailed study  has been hampered by the very short relaxation times that characterize, in conventional systems, the response to an external perturbation,  as well  the inability
to thermally isolate such system from contact with phonon baths or other sources of decoherence, all  leading to the  blurring  of nonequilibrium effects.
The recent advances  follow  the appearance of diverse experimental systems ranging from nano-devices or molecular electronic devices to optically trapped cold atom gases where some of these limitations have been overcome thus allowing a systematic  study of various aspects of non equilibrium response~\cite{blochdali,moritz}. In parallel with the experimental effort much progress was also made on the theoretical front where conceptual ideas and numerical tools have provided new insights into this new and burgeoning field~\cite{polkovnikovrmp,cazalilla,Xiwen,Heid}. These notes are devoted to reviewing a particular direction, the study of {\it quench dynamics in low dimensional quantum systems described by integrable Hamiltonians}.  Both terms require some discussion. In the first part of the Introduction we shall discuss  "quantum quenches" and in the second part "integrable Hamiltonians".

\begin{figure}[htb]
  \centering \includegraphics[width=8cm]{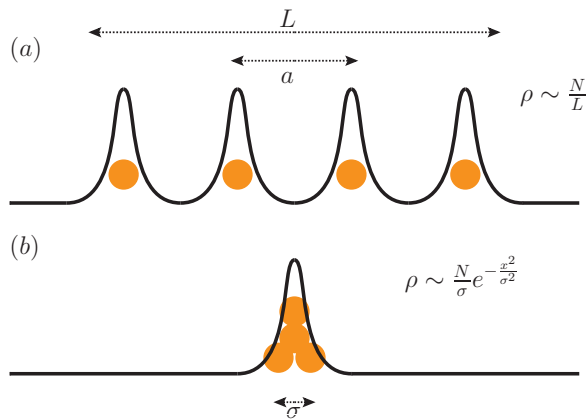}
  \caption{Initial states. $(a)$ For $\frac{a}{\sigma}\gg 1$, we have
    a lattice like state, $\ket{\Psi_{\text{latt}}}$. $(b)$ For $a=0$,
    we have a condensate like state $\ket{\Psi_{\text{cond}}}$,
    $\sigma$ determines the spread.}
  \label{fig:initstate}
\end{figure}

A quantum quench  is  a nonequilibrium process where  a system, initially in some  given quantum state,  is induced to time evolve under the influence of  an applied Hamiltonian. Thus, one prepares the system in some initial state, $|\psi_0\rangle$,  which may in practice be the ground state of an initial Hamiltonian $H_{in}$. The state will have certain characteristic properties and correlations inherited from the initial Hamiltonian. At time $t=0$ one effects some change  and henceforth a new Hamiltonian, $H$, is applied. The state, $|\psi_0\rangle$,  not being an eigenstate of the new Hamiltonian,  will evolve nontrivially  under its influence. 
The response of the system to the quench  reveals its intrinsic time scales, the mechanisms underlying its evolution and the phase transitions it may cross as it evolves.  Many questions arise:  the fate of the initial correlations in the new regime, the development of new correlations, whether the system thermalizes,  among many other.
The changes induced  in the Hamiltonian and leading to the quench can be of a great variety: one may change an interaction coupling constant (e.g via a Feshbach resonance in cold atomic gases), one may release the interacting gas from a harmonic or periodic trap and allow it to expand in open space, one may couple a quantum dot to a Fermi sea and observe the evolution of a Kondo peak.  Further, the quench can be sudden, i.e., over a time window much
shorter than other time scales in the system, it can be driven at a constant
rate or with a time dependent ramp.  Here we shall concentrate on sudden quenches, 
\bea
|\psi,t\rangle = e^{-iHt} |\psi_0\rangle \nn
\eea
where the initial state $\ket{\Psi_{0}}$ describes a system in a finite region
of space with a particular density profile: lattice-like, orF
condensate-like (see Fig.~\ref{fig:initstate})

The quench  dynamics can be studied theoretically and experimentally by observing the time evolution of physical quantities
 that may be local operators,  multi particle  observables,  correlation functions,
local currents or  also global quantities such as entanglements. Consider an observable $\hat{A}$, it will evolve under $H$ as,
\bea
 \langle \hat{A}(t)\rangle= \langle \psi_0| e^{iHt}\hat{A}e^{-iHt} |\psi_0\rangle.  \nn
 \eea
To compute the evolution it is convenient to expand the initial state
in the eigenbasis of the evolution Hamiltonian,
\begin{equation}
  \ket{\Psi_{0}} = \sum_{\{n\}}C_{n}\ket{n},
\end{equation}
where $\ket{n}$ are the eigenstates of $H$ and $C_{n}=
\bra{n}\Psi_{0}\rangle$ are the overlaps with the initial state,
determining the weights with which different eigenstates contribute to
the time evolution:
\begin{equation}\label{eq:quenchoverlaps}
  \ket{\Psi_{0}, t}= \sum_{\{n\}}   e^{-i\epsilon_n t }C_{n}\ket{n}.
\end{equation}
The evolution of observables is then given by,
\bea \label{obs}
    \langle\hat{A}(t)\rangle_{\Psi_0} &= \langle \Psi_{0}, t |\hat{A}\ket{\Psi_{0}, t} = \sum_{\{n, m \}} e^{-i(\epsilon_m-\epsilon_n)t}C^{*}_{n}\,C_m
    \langle n |\hat{A} \ket{m},
\eea
Many issues in dynamics and quantum kinetics revolve around expressions of  this type, some already mentioned: does the system thermalize \cite{rigol2,srednickitherm,deutsch}, equilibrate, cross phase transitions in time, generate
defects as it evolves, what entropy is best suited to describe the evolution, what is the quantum work done in the quench. Some of these questions are discussed in other
lectures of the School and we shall confine our attention to issues related to the dynamics  and how to compute the evolution itself.

\begin{figure}[htb]
  \centering \includegraphics[width=1.5in]{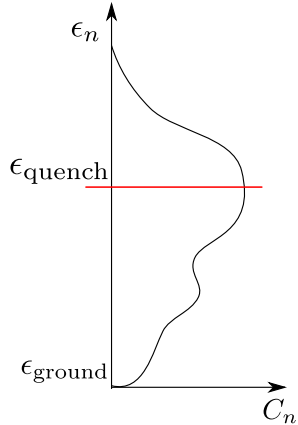}
  \caption{Difference between quench dynamics and
    thermodynamics. After a quench, the system probes high energy
    states and does not necessarily relax to the ground state. In
    thermodynamics, we minimize the energy (or free energy) of a
    system and probe the region near the ground state.}
  \label{fig:energyscales}
\end{figure}

The quench is  typically  not a low energy process. A finite amount of  energy (or energy per unit volume) is  injected into the system,
\begin{equation}
  \epsilon_{\text{quench}} = \bra{\Psi_{0}}H\ket{\Psi_{0}} = \sum_{\{n\}} \epsilon_{n}\abs{C_{n}}^{2},
\end{equation}
and is conserved throughout the evolution. It specifies  the
energy surface on which the system moves (and the temperature, if the system thermalizes.)
This surface is
determined by the initial state through the overlaps $C_n$.  Unlike
the situation in thermodynamics where the ground state and low-lying
excitations play a central role, this is not the case when 
out-of-equilibrium. A quench puts energy into the isolated system which it
 cannot dissipate and  relax to its ground
state. Rather, the eigenstates that contribute to the dynamics depend
strongly on the initial state via the overlaps $C_n$ (see Fig.~\ref{fig:energyscales}).


In the experiments that we seek to describe in this lecture, a system of $N$ bosons is
initially confined to a region of space of size $L$ and then allowed
to evolve on the infinite line while interacting with short range
interactions. The quantum Hamiltonian describing the system in 1-$d$ is called the Lieb-Liniger Hamiltonian,
\bea
  \label{eq:llham}
  H = \int dx \, \partial b^\dag(x) \partial b(x) + c \int  dx \, b^\dag(x)b(x)b^\dag(x)b(x),
\eea
where the field $b^\dagger(x)$ creates a boson at point $x$. The strength  and sign of the coupling constant $c$ can be controlled in the experiment.
When $c > 0$ the system is repulsive,  whereas when $c<0$ the system is attractive and bound states appear. The choice of short range potential $V( x-y)=c\delta(x-y)$
renders the model integrable, as we shall discuss below. 

Several time scales underlie the phenomena we describe.  One is determined by the initial condition (spatial
extent, overlap of nearby wave-functions), and the other is determined by the
parameters of the quenched system (mass, interaction strength).
For an extended system where we start with a locally uniform density
(see Fig.~\ref{fig:initstate}$a$), we expect the dynamics to be in the
constant density regime as long as $t \ll \frac{L}{v}$, $v$ being the
characteristic velocity of propagation. Although the low energy
thermodynamics of a constant density Bose gas can be described by a
Luttinger liquid~\cite{giamarchibk}, we expect the collective
excitations of the quenched system to behave as a highly excited
Liquid since the initial state is far from the ground state. It is
also possible that depending on the energy density
$\epsilon_{\text{quench}}/L$, the Luttinger liquid description may
break down altogether.

\begin{figure}[htb]
  \centering \includegraphics[width=8cm]{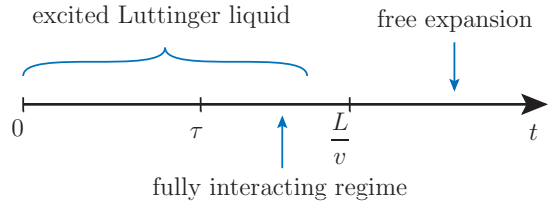}
  \caption{Time scales involved in quench dynamics. $\tau$ is an
    intrinsic time scale that depends on the interaction  strength. $L/v$ is a characteristic time at which the system sees
    the finite extent.}
  \label{fig:timescales}
\end{figure}

The other time scale that enters the description of non equilibrium
dynamics is the interaction time scale, $\tau$, a measure of the time
it takes the interactions to develop fully: $\tau \sim
\frac{1}{c^{2}}$ for the Lieb-Liniger model~\footnote{One can refine
  the estimate for the interaction time setting $\tau \sim
  \frac{1}{\delta E}$, with $\delta E = \displaystyle{\bra{\Phi_0} H_I
    \ket{\Phi_0}}$.  Also if we start from a lattice-like state,
  $\tau$ will include a short time scale $\tau_{a}\sim\frac{a}{v}$
  before which the system only expands as a non-interacting gas, until
  neighboring wave-functions overlap sufficiently.}.  Assuming $L$ is
large enough so that $\tau \ll \frac{L}{v}$, we expect a fully
interacting regime to be operative at times beyond the interactions
scale until $t \sim \frac{L}{v}$  when the density of the system can no
longer be considered constant. In the low density regime  into which the expanding
Lieb-Liniger  gas enters  when $t \gg \frac{L}{v}$ the  effective strong  coupling 
 manifests itself as
fermionization for repulsive interaction and  as bound-state correlations
for  attractive interactions.  Thus the main operation of
the interaction occurs in the time range $\tau \lesssim t \lesssim
\frac{L}{v}$, over which the wave function rearranges and after which
the system is dilute and expands in space while interacting. In this inhomogeneous  low density limit one
can no longer  make contact with thermodynamic ensembles, in particular, with the Generalized
Gibbs ensemble~\cite{rigol3,calabresecardy,cauxessler,cauxkonik}.
When  $L=\infty$, the system is homogenous and free space
expansion is not present. Figure~\ref{fig:timescales} summarizes the
different time-scales involved in a dynamical situation.
We shall also consider initial conditions where the bosons are
``condensed in space'', occupying the same single particle state
characterized by some scale $\sigma$ (see
Fig.~\ref{fig:initstate}$b$).  In this case the short time dynamics is
not present, the time scales at which we can measure the system are
typically much larger than $\frac{\sigma}{N}$ and we expect the
dynamics to be in the strongly interacting and expanding regime.

\hspace*{0.1cm}

We now turn to the second term, "integrable Hamiltonians". It covers  a vast subject
which we shall address only from a particular point of view, the construction of eigenstates
by means of the Bethe Ansatz.
As we saw,  to carry out the computation of the quench dynamics we need to know
the eigenstates of the propagating Hamiltonian. The Bethe Ansatz
approach is helpful in this respect as it provides us with the
eigenstates of a large class of interacting one dimensional
Hamiltonians.  A partial list
includes the Heisenberg chain (magnetism), the Hubbard model
(strong-correlations), the Lieb-Liniger model and Sine-Gordon model (cold atoms in optical
traps), the Kondo model and the Anderson model (impurities in metals,
quantum dots)~\cite{bethe, liebwu,giamarchibk, lieblin, andrei, andreitrieste,
  tsvelik}.  Many of the Hamiltonians that can be thus be solved are
of fundamental importance in condensed matter physics and have been
proposed to describe  various experimental situations. Integrable models   pervade not only condensed matter physics  but appear also in low dimensional quantum and classical field theory and statistical mechanics. They  can be realized experimentally in a variety of ways and many have been extensively studied long before their integrability became manifest.

  For a Hamiltonian to possess   Bethe Ansatz eigenstates  it must have the property that
multi-particle interactions can be consistently factorized into series
of two particle interactions, all of them being
equivalent. Put differently, one must be able to express {\it in a consistent way} multiparticle wave functions in terms of single particle wave functions and 2-particle $S$-matrices.  We shall now  do so for the  Lieb-Liniger Hamiltonian.  The eigenstates we construct will subsequently be used to carry out  quench evolution calculations in the model.
 The derivation of the model and its role in the description of cold atoms experiments are  presented  in other lectures of the School. 
 
 {\bf The Bethe Ansatz eigenstates:}    As the number of bosons is conserved we can consider the Hilbert space of $N$-bosons spanned by states of the form
 \bea
  |F\rangle= \int \prod_{j=1}^N d\,x_j\, F(x_1, \cdots, x_N) \prod_j b^{\dagger}(x_j) |0\rangle, \nn
  \eea
   with $|0\rangle$ being the vacuum state with no bosons, $b(x)|0\rangle=0$. In the $N$-boson sector, when acting on the symmetric wave function $F(x_1, \cdots, x_N)$,  the Hamiltonian takes the form,
\bea
H_N=-\sum_{j=1}^N \frac{\partial^2}{\partial x^2_j} +2c \sum_{j<l} \delta(x_j-x_l). \nn
\eea
The condition that the state $|F\rangle$ be an eigenstate of $H$ is that the wave function $F(x_1, \cdots, x_N)$  satisfy $H_N\,F=E\,F$.

 We proceed to obtain the single particle wave functions and 2-particle $S$-matrices in terms of which we shall express the multiparticle wave functions.  In the 1-boson sector,
$H_1= - \frac{\partial^2}{\partial x^2} $, and the eigenstates are plane waves, $F^\lambda(x)=e^{i\lambda x}$, with energy $E=\lambda^2$.
In the 2-boson sector the interaction term is present,  $H_2=- \frac{\partial^2}{\partial x^2_1} -\frac{\partial^2}{\partial x^2_2}+2c \, \delta(x_1-c_2)$. As the bosons interact only along the boundary, $x_1=x_2$,  we can divide the configuration space into two regions, $x_1<x_2$ and, $x_2<x_1$ in the interior of which the particles are free and where the wave functions take the form:  $Ae^{i(  \lambda_1x_1 + \lambda_2 x_2)}$  and  $Be^{i(  \lambda_1x_1 + \lambda_2 x_2)}$, respectively, with energy: $E= \lambda_1^2 + \lambda_2^2$. The relation between $A$ and $B$ is determined by the interaction which becomes operative on the boundary between the the two regions. The fully symmetrized wave function becomes,
\begin{eqnarray}
F(x_1,x_2)&=& \mathcal{S} e^{i( \lambda_1x_1 +\lambda_2 x_2)}[A \theta(x_2-x_1) +B\theta(x_1-x_2)]=\nn\\
           &=& \; \frac{1}{2} e^{i( \lambda_1x_1 +\lambda_2 x_2)}[A \theta(x_2-x_1) +B\theta(x_1-x_2)] + 
 \frac{1}{2} e^{i( \lambda_1x_2 +\lambda_2 x_1)}[A \theta(x_1-x_2) +B\theta(x_2-x_1)] \nn.
\eea
Here the step function $\theta(x)$ is defined as follows:  $\theta(x)=1$ for $x>0$, $\theta(x)=0$ for $x<0$ and $\theta(x)=\frac{1}{2}$ for $x=0$. Hence $\theta(x)+\theta(-x)=1, \,\forall x$ and $\partial_x \theta(x)= \delta(x)$. 
 Note that the wave function is continuous (as it should be) on the boundary: $F(x,x^+)= F(x,x^-)= \frac{1}{2} e^{i( k_1 +k_2) x}(A+B)$.  
 
 Applying now the Hamiltonian to the wave function, $
 H_2\,F(x_1,x_2)= E\,F(x_1,x_2)$
 one finds that the condition  for $F$ to be an eigenfunction is,
 \bea
 &&-e^{i(\lambda_1+\lambda_2)x_1} \delta(x_1-x_2)[2i\lambda_1(A-B) + 2i\lambda_2(B-A) +2i\lambda_1(A-B)+2i\lambda_2(B-A)]-
 \nn\\
 &&-e^{i(\lambda_1+\lambda_2)x_1} \delta'(x_1-x_2)[(A-B) + (B-A) +(A-B)+(B-A)] +\nn\\
 &&2ce^{i( \lambda_1 +\lambda_2) x_1}\delta(x_1-x_2) \frac{A+B}{2}=0
 \eea
  The $\delta'(x)$ term cancels automatically as result of the continuity
 of the wave function on the boundary.  The requirement that the $\delta(x)$-terms  cancel leads to the condition $
 B= \frac{ (\lambda_1-\lambda_2) +ic}{(\lambda_1-\lambda_2) -ic} A$, from which we deduce the two particle $S$-matrix, 
\bea
S^{12}=\frac{B}{A}= \frac{ (\lambda_1-\lambda_2) +ic}{(\lambda_1-\lambda_2) -ic}, 
\eea
completing  the determination of the two-particle eigenstate:
\bea
  \label{eq:lleigen}
  \ket{\lambda_1 \lambda_2} =\mathcal {N}_{12} \int_x  Z_{12}^{x}(\lambda_1-\lambda_2)  e^{i(\lambda_1
    x_1+\lambda_2x_2)}b^\dag(x_1)b^\dag(x_2)\ket{0}.
\eea
Here $\mathcal {N}_{12}$ is a normalization factor determined by a
particular solution, and
\begin{equation}
  Z^x_{12}(z) = \frac{z-ic\sgn(x_1-x_2)}{z-ic}.
\end{equation}
is the dynamic factor which includes the  $S$-matrix.

The generalization to any number of particles  follows a similar argument. One divides the configuration
space of $N$ bosons into $N!$ regions conveniently  labeled by elements $Q $ of the permutation group $S_N$ which specify their ordering.
Thus the region where $x_{Q1}<x_{Q2} <, \cdots, x_{QN}$  corresponds to the permutation $Q $.  The ordering of the particles in region
$Q$  can be obtained  from the ordering in some standard region, say in region $I$, where $x_1<x_2,\cdots, x_N$, by a series of transpositions $P^{ij}$
where the positions of two adjacent particles    $Qm=i, Q(m+1)=j$  are exchanged.  The amplitude $A_Q$ in region $Q$ is given then as:
\bea
A_Q = \left( \prod S \right) \, A_I
\eea
where the product of $S$-matrices is taken along a path of transpositions leading from $I$ to $Q$. That path is not unique, hence for the construction to be
consistent the S matrices must satisfy the  condition - the Yang-Baxter equation,
\bea
S^{ij}S^{ik}S^{jk}=S^{jk}S^{ik}S^{ij}
\eea
guaranteeing  path independence. This condition is very stringent in general, when particles carry internal degrees of freedom and the $S$ matrices do not commute. Here the condition is satisfied  as the bosons have equal mass hence the individual particle momenta are conserved under scattering and the  $S$-matrices are commuting phases.

The eigenstates  thus take the form 
\bea
|\vec{\lambda} \rangle = \int_x  e^{i \vec{\lambda}\cdot \vec{x}} \sum_Q  A_Q \theta(x_Q) \prod_{j=1}^N b^{\dagger}(x_j)|0\rangle
\eea
where $\int_x \equiv \int \prod_j d\,x_j$ and we omitted the symmetrizer since the wave function is automatically symmetrized under the integration.  The eigenstates may be recast
here into a more convenient form
\bea
  \label{eq:lleigen}
  \ket{\vec{\lambda}} = \mathcal{N}(\vec{\lambda}) \int_x
  \prod_{i<j}Z_{ij}^{x}(\lambda_i-\lambda_j) \prod_j e^{i\lambda_j
    x_j}b^\dag(x_j)\ket{0},
\eea
where $\mathcal{N}(\lambda)$ is a normalization factor determined by a
particular solution.  The energy eigenvalue corresponding to the eigenstate $ \ket{\vec{\lambda}} $ is
\bea \label{en}
E (\vec{\lambda})= \sum_j \; \lambda^2_j.
\eea

The eigesntates and eigenvaues thus obtained play a different role in the study
of equilibrium and  non equilibrium properties of a given quantum system.
To study the thermodynamic properties one
must be able to enumerate and classify all eigenstates  in
order to construct the partition function. To achieve this, some
finite volume boundary conditions (BC) are typically imposed. One may impose
periodic BC to maintain translation invariance or open BC when
the system has physical ends. Imposing periodic boundary conditions  on the Bethe Ansatz wave functions  
of the LL model leads to  $N$ coupled equations,
\bea
e^{i\lambda_j L} = \prod_l \frac{ (\lambda_j-\lambda_l) +ic}{(\lambda_j-\lambda_l) -ic} \quad j=1,\cdots, N.
\eea
whose solutions are the  allowed momentum configurations $\{ \vec{\lambda} \}$ from which the full  spectrum is determined via eq.(\ref{en}).
One can then identify the ground state
and the low lying excitations that dominate the low-temperature
physics.   The analysis and classification of the solutions of the model was given by Lieb-Liniger and by Lieb \cite{lieblin}.  The thermodynamic partition function  was then derived
by summing over all energy eigenvalues with their appropriate degeneracies  by Yang and Yang \cite{Yang2TBA}.

In the study of nonequilibrium  quench dynamics,  on the other hand,   the main issue is
the determination of the eigenstates of the propagating Hamiltonian that enter in the time evolution of a given initial state $|\psi_0\rangle$. This information is encoded 
 the overlaps, $C_n= \langle n|  \psi_0\rangle$,  with the ground state and low lying excitation playing no special role.
  Given the eigenstate expansion of the initial state one needs then to resum  it  with the time phases, eq.(\ref{eq:quenchoverlaps}), to obtain the evolved state.
 The overlaps, however, are not easy  to evaluate due to the complicated nature of
the the Bethe eigenstates and their normalization. The problem is more
pronounced in the far from equilibrium  quench when the state
we start with suddenly finds itself far away from the eigenstates of
the new Hamiltonian [see eq.~\eqref{eq:quenchoverlaps}], and all the
eigenstates have non-trivial weights in the time-evolution.  In all
but the simplest cases, the problem is non-perturbative and the
existing analytical techniques are not suited for a direct application
to such a situation.

When quench calculations are carried out in finite volume all the steps
mentioned are involved: (i) solving the BA equation for the spectrum and the eigenstates
(ii) calculation of overlaps (iii) resummation of the evolution series.  If however one is interested in the physics in the infinite
 volume limit, one need not  (unlike in thermodynamics) pass through finite volume calculation. Instead one can
 carry  out the quench directly in the infinite volume limit allowing the the overlaps to pick out the relevant contributions.
 Working in the infinite volume limit allows us  to replace the discrete sum in eq.(\ref{obs}) by integrations over continuous momentum variables
 transforming the difficult talk of computing overlaps to the simpler one, of calculaeq(ting  residues in Cauchy type integrals, as we discuss below. This approach,
 due to V. Yudson,  circumvents  some of the difficulties mentioned  and leads to an efficient calculational formalism and to transparent physical results.  
 Its application  to the Lieb-Liniger model is the main topic of these notes.  We have applied this approach to other models too, results will be presented elsewhere.

\section{Time evolution on the infinite line}
\label{sec:intro}

In 1985, V.~I.~Yudson presented a new approach to  time evolve
the Dicke model (a model for
superradiance in quantum optics~\cite{dicke}) considered on an infinite line~\cite{yudson1}.
The dynamics in certain cases was
extracted in closed form with much less work than previously required,
and in some cases where it was even impossible with earlier
methods. The core of the method is to bypass the laborious sum over
momenta using an appropriately chosen set of contours and integrating
over momentum variables in the complex plane. It is applicable in its
original form to models with a particular pole structure in the two
particle $S$-matrix, and a linear spectrum. We  generalize the approach to the
case of the quadratic spectrum and apply it to the study of quantum quenches.

As discussed earlier, in order to carry out the quench of a system given at $t=0$  in
a state $|\Psi_0\rangle$ one naturally proceeds by introducing a
``unity'' in terms of a complete set of eigenstates,
\begin{equation}
  \label{eq:timeevol}
    \ket{\Psi_0}
    = \sum_{\{\lambda\}}\ket{\vec{\lambda}}\langle\vec{\lambda}|\Psi_0\rangle
\end{equation}
and then applies  the
evolution operator. The Yudson representation overcomes the difficulties in  computing overlaps and carrying out this sum by
using an integral representation for the (over) complete
basis directly in the infinite volume limit. The argument consists of two independent parts:

{\bf 1.} Since the wave function for bosons, $F(x_1, \cdots, x_N)$, is fully symmetric it suffices to compute overlaps
in a region $(x_1 <,\cdots, <x_N)$. 

 {\bf 2.}  If no boundary conditions are imposed 
then the Schrodinger equation for $N$ bosons,
$H\ket{\vec{\lambda}}=\epsilon(\vec{\lambda})\ket{\vec{\lambda}}$, is
satisfied for any value of the momenta $\{ \lambda_{j}, j=1,\cdots,N
\}$.

 The initial state can then
be written as 
\bea 
\label{Yud1}
 \ket{\Psi_{0}} = \int_{\gamma} d^N \lambda \; C_{\vec{\lambda}}
\; \ket{\vec{\lambda}}
\eea
with the integration over $\vec{\lambda}$
replacing the summation over states. This is akin to summing over an over-complete
basis, the relevant elements in the sum being automatically picked up
by the overlap with the physical initial state. The integration over momenta
will be carried out over contours $\{\gamma\}$, fixed by the pole structure of the $S$-matrices, chosen so 
that eq.(\ref{Yud1}) holds.

We proceed to carry out the evolution
for the repulsive and attractive models for several initial states. The contours of integration,
as we shall see, depend on the sign of the coupling constant $c$.  We will notice that in the repulsive
case, it is sufficient to integrate over the real line.  The attractive case will require the use
of contours separated out in the imaginary direction (to be
qualified below). Those separated contours are consistent with the fact that the spectrum consists of
``strings'' with momenta taking values as complex conjugates. Actually, we shall find that the existence of string solutions (bound states) and their spectrum
follow elegantly and naturally from the contour representation. They need not be  known \`a priori,  or computed from the Bethe Ansatz equations.

\subsection{Repulsive interactions}
\label{sec:repulsive}
We begin by discussing the repulsive case, $c>0$.  For this case, a
similar approach was independently developed in
Ref.~\onlinecite{tracy} and  used by
Lamacraft~\cite{lamacraft} to calculate noise correlations in the
repulsive model. 

 Given a generic  $N$-boson initial state,
\begin{equation}
  \ket{\Psi_{0}} = \int_{\vec{x}} \Phi_s(\vec{x}) \prod_{j} b^{\dag}(x_{j})\ket{0}.
\end{equation}
with $\Phi_s$ symmetrized, it can be rewritten, using the symmetry of the boson
operators, in terms of 
basis states,
\begin{equation}
  \ket{\Psi_{0}} = N!\int_{\vec{x}} \Phi_{s}(\vec x) \ket{\vec{x}}
\end{equation}
where,
\begin{equation}
  \label{eq:initstate}
  \ket{\vec{x}} = \theta(\vec{x}) \prod_j b^\dag(x_j)\ket{0}.
\end{equation}
with $\theta(\vec{x})= \theta(x_1>x_2>\cdots>x_N)$.  It suffices
therefore to show that we can express any coordinate basis state as an
integral over the Bethe Ansatz eigenstates,
\begin{equation}
  \label{eq:yudrep}
  \ket{\vec x} = \theta(\vec{x})\int_\gamma \prod_j\frac{\mathrm{d} \lambda_j}{2\pi} A(\vec{\lambda},\vec{x})\ket{\vec{\lambda}}
\end{equation}
with appropriately chosen contours of integration $\{ \gamma_j \}$ and
$A(\lambda,\vec{x})$, which plays a role similar to the overlap of the
eigenstates and the initial state. Establishing an identity of the type of eq.(\ref{eq:yudrep})  corresponds to identifying the complete
set of eigenstates. That is the reason that the contours for the attractive case, where bound states appear, differ from the repulsive case where they are absent.

We claim that in the repulsive case equation~\eqref{eq:yudrep} is
realized with
\begin{equation}
  A(\vec{\lambda},\vec{x}) = \prod_j e^{-i \lambda_j x_j}
\end{equation}
and the contours $\gamma_j$ running along the real axis from minus to
plus infinity. In other words eqn.~\eqref{eq:yudrep} takes the form,
\bea
  \ket{\vec x} = \theta(\vec{x})\int_{\vec y}\int \prod_{j}\frac{{\rm d}\lambda_{j}}{2\pi}\: \prod_{i<j} Z^y_{ij}(\lambda_i-\lambda_j)
  \times\prod_j e^{i\lambda_j(y_j-x_j)}b^\dag(y_j)\ket{0}.
\eea
Equivalently, we claim that the $\lambda$ integration above produces
$\prod_{j} \delta(y_j-x_j)$.

We shall prove this in two stages. Consider first $y_N>x_N$. To carry
out the integral using the residue theorem, we have to close the
integration contour in $\lambda_N$ in the upper half plane. The poles
in $\lambda_N$ are at $\lambda_j-ic$, $j<N$. These are all below
$\gamma_N$ and so the result is zero. This implies that any non-zero
contribution comes from $y_N\leq x_N$. Let us now consider
$y_{N-1}>x_{N-1}$. The only pole above the contour is $\lambda_{N-1}^*
= \lambda_N+ic$. However, we also have, $y_{N-1}>x_{N-1}>x_{N}\geq y_N
\implies y_{N-1}>y_N$. This causes the only contributing pole to get
canceled. The integral is again zero unless $y_{N-1}\leq x_{N-1}$. We
can proceed is this fashion for the remaining variables thus showing
that the integral is non-zero only for $y_j\leq x_j$.

Now consider $y_1<x_1$. We have to close the contour for $\lambda_1$
below. There are no poles in that region, and the residue is
zero. Thus the integral is non-zero for only $y_1=x_1$. Consider
$y_2<x_2$. The only pole below, at $\lambda_2^*=\lambda_1-ic$ is
canceled as before since we have $y_2<x_2<x_1=y_1$.  Again we get that
the integral is only non-zero for $y_2=x_2$. Carrying this on, we end
up with
\begin{equation}
  \theta(\vec{x})\int_{\vec y} \prod_j \delta(y_j-x_j) b^\dag(y_j)\ket{0} = \ket{\vec x}.
\end{equation}

In order to time evolve this state we act on it with the unitary
time evolution operator. Since the integrals are well-defined we can
move the operator inside the integral signs to obtain,
\begin{equation}
  \ket{\vec x,t} = \theta(\vec{x})\int \prod_{j}\frac{{\rm d}\lambda_{j}}{2\pi}\:
  e^{-i\epsilon(\vec\lambda)t} A(\vec\lambda,\vec x)\ket{\vec{\lambda}}.
\end{equation}

\subsection{Attractive interactions}
\label{sec:attract}
We now consider the case, $c<0$. As mentioned
earlier, the spectrum of the Hamiltonian now contains 
complex, so-called \emph{string solutions} which 
correspond to many-body bound states. In fact, the ground state at
$T=0$ consists of one $N$-particle bound state.  We will see that
Yudson integral representation requires in this case  another set of contours
in order to establish eq.(\ref{eq:yudrep}). Similar
properties are seen to emerge in~\cite{prolhac}, where the authors
obtain a propagator for the attractive Lieb-Liniger model by
analytically continuing the results obtained by Tracy and
Widom~\cite{tracy} for the repulsive model.

One will immediately note that for the attractive interaction the 
structure of the $S$-matrix is altered. This change, $c\to -c$ prevents the proof
of the previous section from working. In particular, the poles in the
variable $\lambda_{N}$ are at $\lambda_{j}+i\abs{c}$ for $j<N$, and
the residue within the contour closed in the upper half plane is not zero
any more. We need choose a contour to avoid this pole. This can be
achieved by separating the contours in the imaginary direction such
that adjacent ${\rm Im}[\lambda_{j}-\lambda_{j-1}] > \abs{c}$. At
first sight, this seems to pose a problem as the quadratic term in
exponent diverges at large positive $\lambda$ and positive imaginary
part.  There are two ways around this. We can tilt the contours as
shown in Fig.~\ref{fig:LLcontours} so that they lie in the convergent
region of the Gaussian integral. The pieces towards the end, that join
the real axes though essential for the proof to work at $t=0$, where
we evaluate the integrals using the residue theorem, do not contribute
at finite time as the integrand vanishes on them as they are taken to
infinity.
\begin{figure}[tb]
  \begin{center}
    \includegraphics[width=8.5cm]{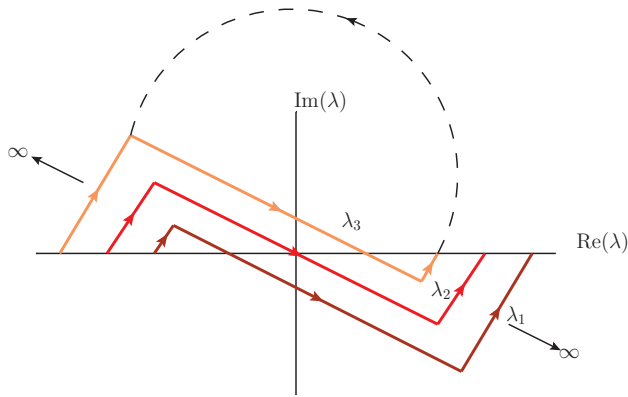}
    \caption{Contours for the $\lambda$ integration. Shown here are
      three contours, and the closing of the $N$th (here, third) contour as
      discussed in the proof.}
    \label{fig:LLcontours}
  \end{center}
\end{figure}
Another more natural means of doing this is to use the finite spatial
support of the initial state. The overlaps of the eigenstates with the
initial state effectively restricts the support for the $\lambda$
integrals, making them convergent.

The proof of equation~\eqref{eq:yudrep} now proceeds as in the
repulsive case. We start by assuming that $y_N>x_N$ requiring us to
close the contour in $\lambda_N$ in the upper half plane. This
encloses no poles due to the choice of contours and the integral is
zero unless $y_N\leq x_N$. Now assume $y_{N-1}>x_{N-1}$.  Closing the
contour above encloses one pole at $\lambda_{N-1}^*=
\lambda_N-i\abs{c}$, however since $y_{N-1}>x_{n-1}>x_N\geq y_N$, this
pole is canceled by the numerator and again we have $y_{N-1}\leq
x_{N-1}$. We proceed in this fashion and then backwards to show that
the integral is non-zero only when all the poles cancel, giving us
$\prod_j \delta(y_j-x_j)$, as required.

\subsection{Two particle dynamics}
\label{sec:2pquench}
We begin with a detailed discussion of the quench dynamics of two
bosons. As we saw, it is convenient to express any initial state in
terms of an ordered coordinate basis, $ \ket{\vec{x}} =
\theta(x_1>x_2>\cdots>x_N)\prod_j b^\dag(x_j)\ket{0}$.  At finite
time, the wave function of bosons initially localized at $x_1$ and
$x_2$ and subsequently evolved by a repulsive Lieb-Liniger Hamiltonian
is given by,
\bea
  \label{eq:rep2p}
    \ket{\vec{x},t}_2 &=& e^{-iHt}\theta(x_1-x_2) b^{\dagger}(x_1) b^{\dagger}(x_2) \ket{0}  = \nn \\
    &&=\int_{y,\lambda} Z^y_{12}(\lambda_1-\lambda_2)  e^{-i\lambda_1^2t-i\lambda_2^2t+i\lambda_1(y_1-x_1)+i\lambda_2(y_2-x_2)}  b^\dag(y_1)b^\dag(y_2)\ket{0}\\
    &&= \int_y \frac{e^{i\frac{(y_1-x_1)^2}{4t}+i\frac{(y_2-x_2)^2}{4t}}}{4\pi i t}
    \times\Big[1- c\sqrt{\pi i
      t}\theta(y_2-y_1)e^{\frac{i}{8t}\alpha^2}
    \erfc\left(\frac{i-1}{4}\frac{i\alpha}{\sqrt{t}}
    \right)\Big]  b^\dag(y_1)b^\dag(y_2)\ket{0} \nn
\eea
where $\alpha = 2ct-i(y_1-x_1)-i(y_2-x_2)$.  The above expression
retains the Bethe form of wave functions defined in different
configuration sectors. The only scales in the problem are the
interaction strength $c$ and $x_1-x_2$,  the initial
separation between the particles.

In order to get physically meaningful results we need to start from a
physical initial state. We choose the state
$\ket{\Psi(0)_{\text{latt}}}$ where bosons are trapped in a periodic
trap forming initially a lattice-like state (see
fig.~\ref{fig:initstate}a),
\begin{equation}
  \label{eq:harmtrap}
  \ket{\Psi(0)_{\text{latt}}} = \prod_j   \left[ \frac{1}{(\pi\sigma^2)^{\frac14} } \int_{\vec{x}}e^{-\frac{(x_j+(j-1)a)^2}{2\sigma^2}}   b^\dag(x_{j})\right]\ket{0}.
\end{equation}
We shall consider two limits. The first, to which we continue to refer to as $\ket{\Psi(0)_{\text{latt}}}$, is  $\sigma \ll a$,
with wave functions of neighboring bosons  not
overlapping significantly, i.e., $ e^{-\frac{a^2}{\sigma^2}} \ll 1$. In this case 
the ordering of the initial particles needed for the Yudson
representation is induced by the non-overlapping support and it
becomes possible to carry out the integral analytically. The other limit, $a \ll \sigma$, when there is maximal initial overlap will be referred to as  
condensate (in position space) wave function $| \Psi(0)_{\text{cond}}\rangle$.

{\bf Evolution of the density:}   We consider the evolution of density   $\hat{\rho}(x)=b^{\dag}(x)b(x)$ at $x=0$ in the  state
$\ket{\Psi(0)_{\text{latt}}}$. Fig.~\ref{fig:twores}
shows $\bra{\Psi_{\text{latt}} ,t} \rho(0)\ket{\Psi_{\text{latt}},t}$
for repulsive, attractive and non-interacting bosons. No difference is
discernible between the three cases. The reason is obvious: the local
interaction is operative only when the wave functions of the particles
overlap. As we have taken $\sigma \ll a$ this will occur only after a
long time when the wave-function is spread out and overlap is
negligible.
\begin{figure}[htb]
  \centering \includegraphics[width=8.7cm]{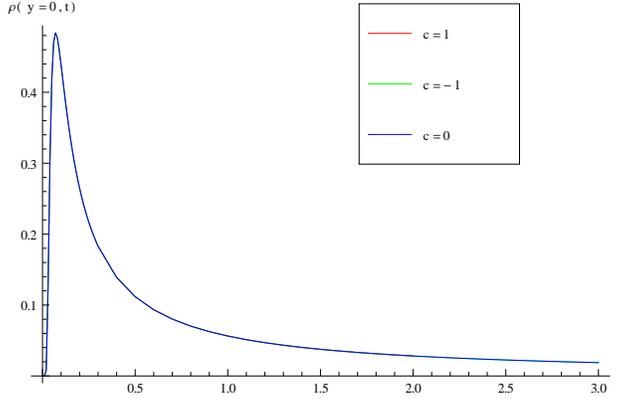}
  \caption{(Color online) $\langle\rho(x=0,t)\rangle$ vs. $t$, after
    the quench from $\ket{\Psi_{\text{latt}}}$. $\sigma/a\sim
    0.1$. The curves appear indistinguishable (i.e. lie on top of each other) since the particles
    start out with non significant overlap. The interaction effects
    would show up only when they have propagated long enough to have spread
    sufficiently to reach a significant overlap, at which time the
    density is too low.}
  \label{fig:twores}
\end{figure}

When the separation $a$ is set to zero, with maximal initial overlap between the bosons, $|
\Psi(0)_{\text{cond}}\rangle$, we expect the interaction to be operative from the start.
Fig.~\ref{fig:tworescond} shows the density
evolution for attractive, repulsive and no interaction.  The decay of
the density is slower for attractive model than the for the
non-interacting which in turn is slower than for the repulsive model.
\begin{figure}[htb]
  \centering \includegraphics[width=8.5cm]{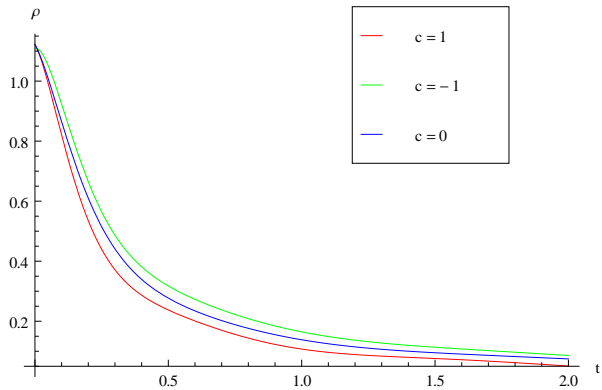}
  \caption{(Color online) $\langle\rho(x=0,t)\rangle$ vs. $t$, after
    the quench from $\ket{\Psi_{\text{cond}}}$. $\sigma \sim 0.5,\:
    a=0$. As the bosons overlap interaction effects show up
    immediately. Lower line: $c=1$, Upper line: $c=-1$, Middle line: $c=0$. }
  \label{fig:tworescond}
\end{figure}
Still, the density does not show much difference between repulsive and
attractive interactions in this case.  A drastic difference
will appear when we study the noise correlations $\langle \Psi_0,t
|\rho(x_1)\rho(x_2)|\Psi_0,t \rangle$, as will be shown below. 

{\bf The appearance of bound states:}  Actually the apparent similarity in the behavior observed
in Fig.~\ref{fig:tworescond} is somewhat misleading. We shall show here that for attractive interaction
bound states appear in the spectrum, though  their effect in evolution - in the case of two bosons - can be absorbed
into the same mathematical expression as for the repulsive case.

 Recall that the
contours of integration are separated in the imaginary direction.  In
order to carry out the integration over $\lambda$, we shift the
contour for $\lambda_{2}$ to the real axis, and add the residue of the
pole at $\lambda_{2}=\lambda_{1}+i\abs{c}$.  The two particle finite
time state can be written as
\bea   &&\ket{\vec{x},t}_{2}
    =\int_y\int_{\gamma_c}\prod_{i<j=1,2}Z^y_{ij}(\lambda_i-\lambda_j) \prod_{j=1,2}
    e^{-i\lambda_{j}^{2}t+i\lambda_j (y_{j}-x_j)}b^\dagger(y_j)\ket{0}\\
    &&=\int_y\bigg[\int_{\gamma_r}\prod_{i<j,1}^2Z^y_{ij}(\lambda_i-\lambda_j)\prod_{j,1}^2
    e^{-i\lambda_{j}^{2}t+i\lambda_j(y_{j} -x_j)}b^\dagger(y_j)\ket{0}+ \theta(y_{2}-y_{1})I(\lambda_2=\lambda_1+i\abs{c},t)
    b^\dagger(y_1)b^\dagger(y_2)|0\rangle\bigg] \nn
\eea
$\gamma_c$ refers to contours that are separated in imaginary
direction, $\gamma_r$ refers to all $\lambda$ integrated along real
axis. $I(\lambda_2=\lambda_1+i \abs{c},t)$ is the the residue obtained
by shifting the $\lambda_{2}$ contour to the real axis from the pole
at $\lambda_2$ at $\lambda_1+i\abs{c}$. It is given by
\bea
  \label{eq:att2pbd}
    &&I(\lambda_2=\lambda_1+i \abs{c},t) = -2c\int_y\int_{\lambda_1}e^{i\lambda_1(y_1-x_1)+i(\lambda_{1}+i\abs{c})(y_2-x_2)-i\lambda_{1}^{2}t - i(\lambda_{1}+i\abs{c})^{2}t}\\
    &&=-2c\int_y\int_{\lambda_1}e^{i\lambda_1(y_1-x_1+y_2-x_2)-\frac{\abs{c}}{2}(y_2-y_1)-\frac{\abs{c}}{2}(x_1-x_2)} e^{-i(\lambda_{1}-i\abs{c}/2)^{2}t - i(\lambda_{1}+i\abs{c}/2)^{2}t}  \nn \\
    &&=-c\int_y\int_{\lambda_1}e^{i\lambda_1(y_1-x_1+y_2-x_2)-\frac{\abs{c}}{2}|y_2-y_1|-\frac{\abs{c}}{2}(x_1-x_2)} e^{-2i\lambda_{1}^{2}t + i\frac{\abs{c}^{2}}{2}t} \nn.
\eea
 This second term corresponds
to the two-particles propagating as a bound state with the  wave function  ($e^{-\frac{\abs{c}}{2}|y_2-y_1|}$) and with center of mass position and momentum ($e^{i\lambda_1(y_1+y_2-x_2-x_{1})}$). It has  a kinetic energy $2\lambda_{1}^{2}$ and  a binding energy of $-c^{2}$, as reported in~\cite{yang}.  Also the overlap of the bound state with the initial state follows immediately from the representation. It is given by $-ce^{-\frac{\abs{c}}{2}(x_1-x_2)} $
Such bound states appear for any
number of particles involved.  For instance, for three particles, the
Yudson representation with complex $\lambda$'s automatically produces
multiple bound-states coming from the poles, i.e. $
I(\lambda_2=\lambda_1+i\abs{c})$, $I(\lambda_3=\lambda_1+i\abs{c})$,
etc. They give rise to two and three particle bound states of the form
$e^{-\frac{|c|}{2}(|y_1-y_2|+|y_1-y_3|+|y_2-y_3|)}$. The contribution
to the above integral coming from the real axis corresponds to the
non-bound state.

Finally, putting all together
\bea
  \label{eq:att2p}
    &&\ket{\vec{x},t}_2
    = \int_y \frac{e^{i\frac{(y_1-x_1)^2}{4t}+i\frac{(y_2-x_2)^2}{4t}}}{4\pi i t} \left[1+ \abs{c}\sqrt{\pi i
        t}\theta(y_2-y_1)e^{\frac{i}{8t}\tilde\alpha^2}\erfc\left(\frac{i-1}{4}\frac{i\tilde\alpha}{\sqrt{t}}
      \right)\right] b^\dagger(y_1)b^\dagger(y_2)|0\rangle \nn
\eea
where $\tilde\alpha = -2\abs{c}t-i(y_1-x_1)-i(y_2-x_2)$. Surprisingly,
the wave function maintains its form and we only need to replace $c
\to -c$.  This simple result is not valid for more than two particles.

{\bf Evolution of  noise correlations:}    The density-density correlation function, $\langle \Psi_0,t
|\rho(x_1)\rho(x_2)|\Psi_0,t \rangle$, involves 
the interaction in an essential way  as the geometry of the experimental 
set-up measures the interference of ``direct'' and ``crossed''
propagating waves, see fig.~\ref{fig:hbt}a.
\begin{figure}[hbt]
  \centering \includegraphics[width=7.5cm]{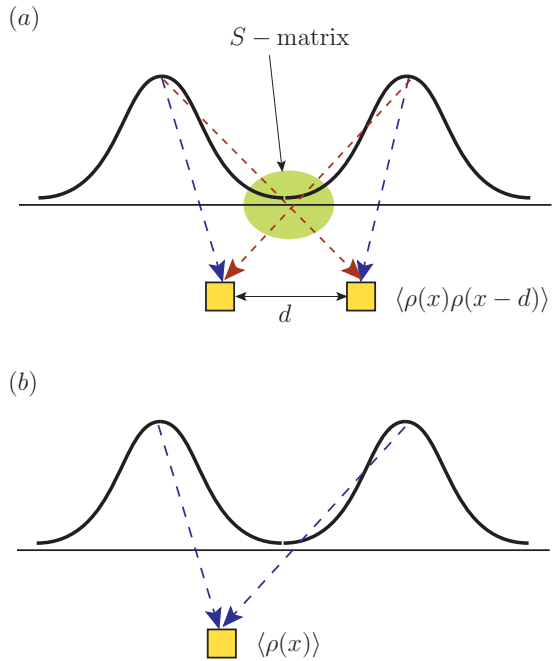}
  \caption{$(a)$  The Hanbury-Brown Twiss effect, where
    two detectors are used to measure the interference of the direct (big dashes)
    and the crossed waves (small dashes). The $S$-matrix enters explicitly. $(b)$ The
    density measurement is not directly sensitive to the
    $S$-matrix. The thick black line shows the wave-function
    amplitude, the dotted lines show time propagation.}
  \label{fig:hbt}
\end{figure}
The interaction among bosons is expected therefore to have  a  significant effect in noise measurements, more so than  in  density measurements which do not directly involve scattering,  see fig.~\ref{fig:hbt}$b$.

The set-up  is the famous Hanbury-Brown Twiss experiment~\cite{hbt} where for
free bosons or fermions, the crossing produces a phase of $\pm 1$ and
causes destructive or constructive interference. Originally designed to study the 
photons arriving from two stars providing  constant sources of light, in our case the 
set-up is generalized to multiple time dependent sources with the phase
given by the two particle $S$-matrix capturing the interactions
between the particles.  In Fig.~\ref{fig:corr2platt} we present the
density-density correlation matrix $\langle \rho(x_1)\rho(x_2)\rangle$ for
the repulsive gas, attractive gas, and the non-interacting gas, shown
at different times, starting with the lattice initial
state of two bosons. Figure~\ref{fig:corr2pcond} shows the same for the condensate
initial state.
\begin{figure}[htb]
  \centering \includegraphics[width=7cm]{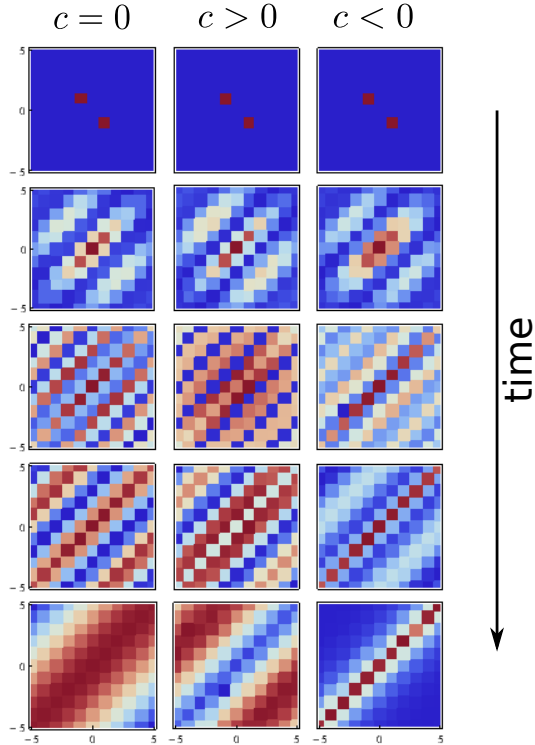}
  \caption{(Color online) Time evolution of density-density
    correlation matrix ($\langle \rho(x)\rho(y)\rangle$) for the
    $\ket{\Psi_{\text{latt}}}$ initial state.  Blue is zero and red is
    positive. The repulsive model shows anti-bunching, i.e.,
    fermionization at long times, while the attractive model shows
    bunching.}
  \label{fig:corr2platt}
\end{figure}
\begin{figure}[htb]
  \centering \includegraphics[width=7cm]{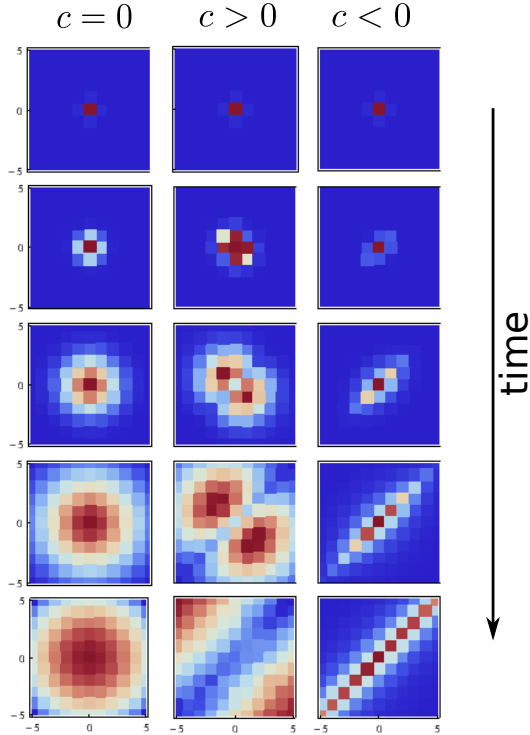}
  \caption{(Color online) Time evolution of density-density
    correlation matrix ($\langle \rho(x)\rho(y)\rangle$) for the
    $\ket{\Psi_{\text{cond}}}$ initial state.  Blue is zero and red is
    positive. The repulsive model shows anti-bunching, i.e.,
    fermionization at long times, while the attractive model shows
    bunching.}
  \label{fig:corr2pcond}
\end{figure}
In both initial states we note that the repulsive gas develops  strong anti-bunching, repulsive correlations at long times akin to fermonic correlations
while  the attractive gas shows strong bunching correlations, enhanced bosonic in nature.  We shall discuss these in detail below.
We expect the results to be qualitatively similar for higher particle
number, though beyond two particles the integrations
cannot be carried out exactly. However, we can extract the asymptotic
behavior of the wave functions analytically, as we show below.

\subsection{Multiparticle dynamics at long times}
\label{sec:asymp}

Here we derive expressions for  multiparticle
wavefunction evolution at long times.  The number of particles $N$ is
kept fixed in the limiting process, hence, as discussed in the
Introduction in the long time limit  we are in the low density limit where interactions are
expected to be dominant.  The other regime where $N$ is sent to
infinity first will be discussed in a separate report.  We first deal
with the repulsive model, for which  no bound states exist and
the momentum integrations can be carried out over the real line  and
then proceed to the attractive model. In a separate sub-section, we
examine the effect of starting with a condensate-like initial state.

\hspace*{.1cm}

{\bf Repulsive interactions - quench asymptotics in the lattice state:}  At
large time, we use the stationary phase approximation to carry out the
$\lambda$ integrations.  The phase oscillations come primarily from
the exponent $e^{-i\lambda_{j}^{2}t +i\lambda_{j}(y_{j}-x_{j})}$.  At
large $t$ (i.e., $t\gg\frac{1}{c^{2}}$), the oscillations are rapid,
and the stationary point is obtained by solving
\begin{equation}
  \frac{\rmd}{\rmd \lambda_{j}} [-i\lambda_{j}^{2}t + i\lambda_{j}(y_{j}-x_{j})]= 0.
\end{equation}
Note that typically one would ignore the second term above since it
doesn't oscillate faster with increasing $t$, but here we cannot since
the integral over $y$ produces a non-zero contribution for $y\sim t$
at large time.  Doing the Gaussian integral around this point (and
fixing the $S$-matrix prefactor to its stationary value), we obtain
for the repulsive case,
\bea
  \label{eq:inftime}
  \ket{\vec{x},t}  \to  \int_y \prod_{i<j} Z^y_{ij}\left(\frac{y_i-y_j-x_i+x_j}{2t}\right)
 \prod_j \frac{1}{\sqrt{4\pi
      it}}e^{-i\frac{(y_j-x_j)^2}{4t}+i\frac{(y_j-x_j)^2}{2t}}
  b^\dag(y_j)\ket{0}.
\eea
In the above expression, the wavefunction has support mainly from
regions where $y_j/t$ is of order one.  In an experimental setup, one
typically starts with a local finite density gas, i.e., a finite
number of particles localized over a finite length. With this
condition, at long time, we can neglect $x_{j}/t$ in comparison with
$y_{j}/t$, giving
\begin{equation}
  \label{eq:inftime2}
  \ket{\vec{x},t} \to  \int_y \prod_{i<j} Z_{ij}(\xi_i-\xi_j)
  \prod_j \frac{1}{\sqrt{4\pi it}}e^{it\xi_j^2-i\xi_j x_j} b^\dag(y_j)\ket{0}
\end{equation}
where $\xi=\frac{y}{2t}$.

We turn now to calculated the asymptotic evolution of some
observables.  To compute the expectation value of the density we start
from the coordinate basis states, $\bra{\vec{x'},t
}\rho(z)\ket{\vec{x},t}$ which we then integrate with the chosen
initial state,
\bea
  \bra{\vec{x'},t }\rho(z)\ket{\vec{x},t} =
  \sum_{\{P\}}\int_y \sum_j\delta(y_j-z)
  \prod_{i<j} Z_{ij}(\xi_i-\xi_j)Z^*_{P_i
    P_j}(\xi_{P_i}-\xi_{P_j}) \prod_j \frac{1}{4\pi t}e^{-i(\xi_j
    x_j-\xi_{P_j}x'_j)} \nn
\eea

Note that the above product of $S$-matrices is actually independent of
the ordering of the $y$. First, only those terms appear in the product
for which the permutation $P$ has an inversion. For example, say for
three particles, if $P=312$, then the inversions are 13 and 23. It is
only these terms which give a non-trivial $S$-matrix contribution.
For the non-inverted terms, here 12, we get
\begin{equation}
  \frac{\xi_1-\xi_2-ic\sgn(y_1-y_2)}{\xi_1-\xi_2-ic} \frac{\xi_1-\xi_2+ic\sgn(y_1-y_2)}{\xi_1-\xi_2+ic}
\end{equation}
which is always unity irrespective of the ordering of $y_1,y_2$. For a
term with an inversion, say 23, we get,
\begin{equation}
  \frac{\xi_2-\xi_3-ic\sgn(y_2-y_3)}{\xi_2-\xi_3-ic} \frac{\xi_3-\xi_2+ic\sgn(y_3-y_2)}{\xi_3-\xi_2+ic}
\end{equation}
which is always equal to
\begin{equation}
  \frac{\xi_2-\xi_3+ic}{\xi_2-\xi_3-ic}\equiv S(\xi_2-\xi_3)
\end{equation}
irrespective of the sign of $y_2-y_3$. This allows us to carry out the
integration over the $y_j$.

In order to calculate physical observables, we have to choose initial
states.  We treat here the lattice state
with $N$ particles distributed  uniformly in a series of
harmonic traps given by,
\begin{equation}
  \label{eq:init1}
  \ket{\Psi_{\text{latt}}} = \int_x \prod_{j=1}^N \frac{1}{(\pi \sigma^2)^{\frac14}}e^{-\frac{(x_j+(j-1)a)^2}{2\sigma^2}}b^\dag(x_j)\ket{0},
\end{equation}
such that the overlap between the wave functions of two neighboring
particles is negligible. In this particular case, the ordering of the
particles is induced by the limited non-overlapping support of the
wave function.

In this  lattice-like state the initial wave function starts out with
the neighboring particles having negligible overlap. At short time (as
seen from \eqref{eq:rep2p}), the particle repel each other and
never cross due to the repulsive interaction. So at long  time, the
interaction does not play a role since the wave functions are
sufficiently non-overlapping. It is only the $P=1$ contribution then
that survives, and we get for the density
\begin{equation}
  \bra{\vec{x}',t}\rho(z)\ket{\vec{x},t} =
  \sum_j \prod_{k\ne j}\frac{\delta(x_k-x_k')e^{-i \frac{z}{2t}(x_j-x_j')}}{4\pi t}.
\end{equation}
We need to integrate the position basis vectors $\ket{\vec{x}}$ over
some initial condition. We do this here for the lattice
state~\eqref{eq:init1} This gives
\begin{equation}
\rho_{\text{latt}}(\xi_z) =  \bra{\Psi_{\text{latt}},t} \rho(z) \ket{\Psi_{\text{latt}},t}=\ \frac{N\sigma}{2\sqrt{\pi}t}e^{-\frac{\xi_z^2}{\sigma^2}}
\end{equation}

Mathematically, any $S$-matrix factor that appears will necessarily
have zero contribution from the pole - this is easy to see from the
pole structure, and the ordering of the coordinates.  In order to get
a non-zero result, we need to fix at least two integration variables
(i.e., the $y_{j}$). Thus the first non-trivial contribution comes
from the two-point correlation function.

We now proceed to calculate the evolution of the noise, i.e., the two
body correlation function 
\bea
\rho_2(z,z'; t)_{\text{latt}} =
\bra{\Psi_{\text{latt}},t}\rho(z)\rho(z') \ket{\Psi_{\text{latt}},t}
\eea
 The contributions can
be grouped in terms of number of crossings, which corresponds to
a grouping in terms of the coefficient $e^{-ca}$~\cite{lamacraft}. The leading order
term can be explicitly evaluated and we show below which terms
contribute. In general we have
\bea
  \rho_{2\text{ latt}}(z,z';t) = \sum_{\{P\}}\int_y\left(\sum_{j,K}\delta(y_j-z)\delta(y_k-z')\right) \prod_{i<j,(ij)\in P} S(\xi_i-\xi_j) \prod_j \frac{1}{4\pi
    t}e^{-i(\xi_j x_j-\xi_{P_j}x'_j)}. \nn
\eea
The above shorthand in the $S$-matrix product means that only the
$(ij)$ that belong to the inversions in $P$ are included.  A detailed  discussion
of how to carry out the summation is given in ref. I. Here we quote the result,

\bea
  \rho_{2 \text{ latt}}(z,z') = \frac{N^2\sigma^2}{4\pi
    t^2}e^{-(\xi_x^2+\xi_{z'}^2)\sigma^2}\bigg[1+
  \frac{2}{N^2}\mathrm{Re}S(\xi_{z}-\xi_{z'})  e^{ia(z-z')}
  \frac{N(1-e^{ia(z-z')}g)+e^{iaN(z-z')}g_{zz'}^N-1}{[1-g_{zz'}e^{ia(z-z')}]^2}\bigg] \nn
\eea
where
\bea
  g_{zz'} =
  1-2c\sqrt{\pi}\sigma S(\xi_z-\xi_{z'}-ic) \Big[e^{(c+i\xi_z)^2\sigma^2}\erfc\{(c+i\xi_z)\sigma\}
  +e^{(c-i\xi_{z'})^2\sigma^2}\erfc\{(c-i\xi_{z'})\sigma\}\Big] \nn
\eea

To compare with the Hanbury-Brown Twiss result, we calculate the
normalized spatial noise correlations, given by $C_2(z,z')\equiv
\frac{\rho_2(z,z')}{\rho(z)\rho(z')} - 1$.  In the
non-interacting case, i.e., $c=0$, $S(\xi)=1$ and $g_{zz'}=0$ and we
recover the HBT result for $N=2$,
\begin{equation}
  C_2^0(\xi_z,\xi_{z'}) = \frac12 \cos(a(\xi_z-\xi_z'))
\end{equation}
One can also check that the limit of $c\to\infty$ gives the expected
answer for free fermions, namely,
\begin{equation}
  C_2^\infty(\xi_z,\xi_{z'}) = -\frac12 \cos(a(\xi_z-\xi_z'))
\end{equation}
At finite $c$ we can see a sharp fermionic character appear that
broadens with increasing $c$ as shown in Fig.~\ref{fig:largec}.
\begin{figure}[h]
  \centering \includegraphics[width=8cm]{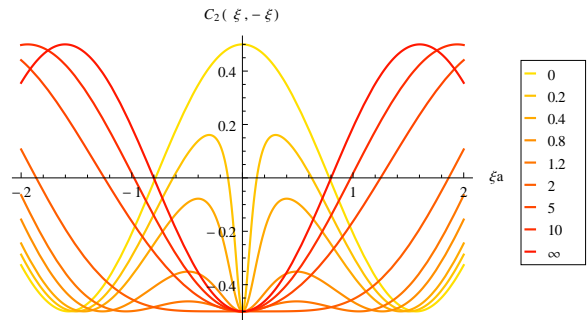}
  \caption{(Color online) Normalized noise correlation function
    $C_2(\xi,-\xi)$.  Fermionic correlations develop on a time scale
    $\tau\sim c^{-2}$, so that for any $c$ we get a sharp fermionic
    peak near $\xi=0$, i.e., at large time.  The key shows values of
    $ca$ (from Ref.~\onlinecite{deeprl}).}
  \label{fig:largec}
\end{figure}
The large time behavior is captured in a small window around
$\xi=0$. One can see that at any finite $c$, the region near zero
develops a strong fermionic character, thus indicating that
irrespective of the value of the coupling that we start with, the
model flows towards an infinitely repulsive model at large time, that
can be described in terms of free fermions. We also obtained this
result ``at'' $t=\infty$ at the beginning of this section.

For higher particle number, we see ``interference fringes''
corresponding to the number of particles, that get narrower and more
numerous with an increase, memory of the initial lattice
state. However, the asymptotic fermionic character does not
disappear. Figures \ref{fig:5prep} and \ref{fig:10prep} show the noise
correlation function for five and ten particles respectively. The
large peaks are interspersed by smaller peaks and so on. This reflects
the character of the initial state.
\begin{figure}[htb]
  \centering \includegraphics[width=2.5in]{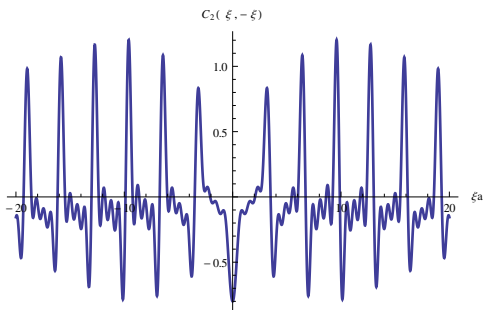}
  \caption{Normalized noise correlation function for five particles
    released for a Mott-like state for $c>0$ (from
    Ref.~\onlinecite{deeprl})}
  \label{fig:5prep}
\end{figure}
\begin{figure}[htb]
  \centering \includegraphics[width=2.5in]{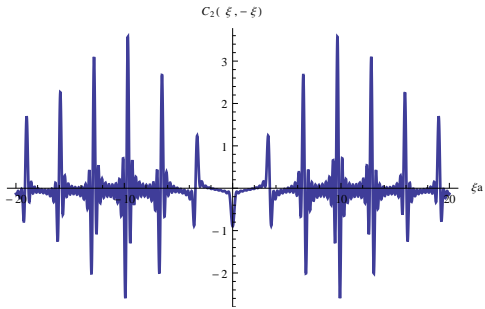}
  \caption{Normalized noise correlation function for ten particles
    released for a Mott-like state for $c>0$.}
  \label{fig:10prep}
\end{figure}

{\bf Attractive interactions - quench asymptotics in the lattice state:}
For the attractive case, since the contours of integration are spread
out in the imaginary direction, we have the contributions from the
poles in addition to the stationary phase contributions at large time.
The stationary phase contribution is picked up on the real line, but
as we move the contour, it stays pinned above the poles and we need to
include the residue obtained from going around them, leading to sum
over several terms.Fig.~\ref{fig:attasymp} shows an example of how
this works.
\begin{figure}[htb]
  \centering \includegraphics[width=6cm]{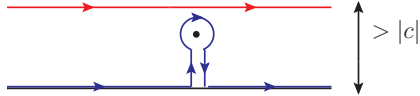}
  \caption{(Color online) Contribution from stationary phase and pole at large time
    in the attractive model. The blue (lower) contour represents the shifted
    contour.}
  \label{fig:attasymp}
\end{figure}

In Ref.~\onlinecite{deeprl}, a formula was provided for the asymptotic
state.  Here we give a more careful treatment by taking into account
that the fixed point of the approximation moves for terms that come
from a pole of the $S$-matrix. It is therefore necessary to first
shift the contours of integration, and then carry out the integral at
long time. We carry this out below.

Shifting a contour over a pole leads to an additional term from the
residue:
\begin{equation}
  \int_{\gamma_{2}}\frac{{\rm d}\lambda_{2}}{2\pi}\to \int_{\gamma^{\rm R}_{2}} \frac{{\rm d}\lambda_{2}}{2\pi} - i\mathcal{R}(\lambda_{2}\to\lambda_{1}+i\abs{c})
\end{equation}
where $\mathcal{R}(x)$ indicates that we evaluate the residue given by
the pole $x$. $\gamma_{j}$ indicates the original contour of
integration and $\gamma_{j}^{\rm R}$ indicates that integration is
carried out over the real axis.  Proceeding with the other variables
we end up with
\begin{multline} \label{eq:attasymp}
  \int_{\gamma_{1},\gamma_{2},\cdots,\gamma_{N}}\to \int_{\gamma^{\rm
      R}_{1}} \left[\int_{\gamma^{\rm R}_{2}} +
    i\mathcal{R}(\lambda_{2}\to\lambda_{1}+i\abs{c})\right]
  \left[\int_{\gamma^{\rm R}_{3}} + i\mathcal{R}(\lambda_{3}\to\lambda_{1}+i\abs{c}) + i\mathcal{R}(\lambda_{3}\to\lambda_{2}+i\abs{c})\right]\cdots\\
  \times\Big[\int_{\gamma^{\rm R}_{N}} +
  i\mathcal{R}(\lambda_{N}\to\lambda_{1}+i\abs{c}) +
  i\mathcal{R}(\lambda_{N}\to\lambda_{2}+i\abs{c}) + 
   \cdots + i\mathcal{R}(\lambda_{N}\to\lambda_{N-1}+i\abs{c}) \Big] \nn
\end{multline}
The integrals can now be evaluated using the stationary phase
approximation. The correction produced by the above procedure does not
affect the qualitative features observed in Ref.~\onlinecite{deeprl}.

We now calculate the evolution of the density and the two body
correlation function in order to compare with the repulsive case.  We
will first study the two particle case. Although we have a finite time
expression for this case from which we can directly take a long time
limit, we will study the asymptotics using the above scheme for an
$N$-particle state, since we have an analytical expression to go with.
We get two terms, the first being the stationary phase contribution,
and is just like the repulsive case with $c\to-c$. The second is the
contribution from the pole. It contains the bound state contribution
which brings about another interesting feature of the attractive
case. While the asymptotic dynamics of the repulsive model is solely
dictated by the new variables $\xi_j\equiv\frac{y_j}{2t}$, and all the
time dependence of the wave function enters through this ``velocity''
variable, this is not the case in the attractive model. While it is
true that the system is naturally described in terms of $\xi$
variables, there still exists non-trivial time dependence.

First, we integrate out the $x$ dependence assuming an initial
lattice-like state. This gives,
\bea
  \label{eq:attasympnox}
  \ket{\Psi_{\text{latt}}(t)} = \int_y
  \sum_{\xi_j^*=\xi_j,\xi_i^*+ic,i<j}
  \prod_{i<j} S_{ij}(\xi_i^*-\xi_j^*)\prod_j \frac{(4\pi \sigma^2)^{\frac14}}{\sqrt{4\pi it}} e^{-(\sigma^2/2+it)(\xi_j^*)^2+i\xi_j^*(2t\xi_j+a(j-1))}
  b^\dag(y_j)\ket{0}.\nn
\eea
Defining $\phi(\xi,t)$ from $\ket{\Psi_{\text{latt}}(t)} = \int_y
\phi(\xi,t)\prod_j b^\dag(y_j)\ket{0}$, we have for the density
evolution under attractive interactions, $c<0$,
\begin{equation}
  \label{eq:densatt}
  \rho_{\text{latt}}^-(z;t) = \sum_{\{P\},j}\int_y \delta(y_j-z)\phi^*(\xi_P,t)\phi(\xi,t)
\end{equation}
We can show numerically (the expressions are a bit unwieldy to write
here), that asymptotically, the density shows the same Gaussian
profile that we expect from a uniformly diffusing gas, namely,
$e^{-\xi^2 \sigma^2}$.

With this, we can proceed to compute the noise correlation
function. The two particle case is easy, as there are no more
integrations to carry out. We get,
\begin{equation}\begin{split}
    \label{eq:noiseatt}
    \rho_{2\text{ latt}}^-(z,z';t) = \sum_{\{P\},j,k}\int_y \delta(y_j-z)\delta(y_k-z')\phi^*(\xi_P,t)\phi(\xi,t)= \lvert\phi_s(\xi_z,\xi_z')\rvert^2,
  \end{split}
\end{equation}
where $\phi_s$ is the symmetrized wavefunction.
Fig.\ref{fig:c2atttime} shows the normalized noise correlations for
different values of $t$.
\begin{figure}[h]
  \centering
  \includegraphics[width=8.5cm]{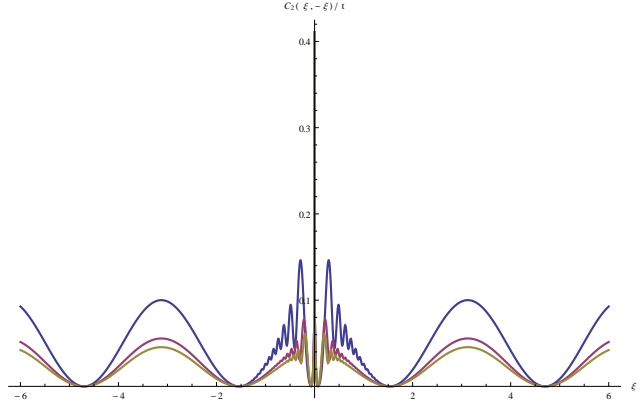}
  \caption{(Color online) Variation of $C_2$ for the attractive case
    with time. Note the growth of the central peak. At larger
    times, the correlations away from zero fall off. $ta^2=20,40,60$
    for blue (top), magenta (middle) and yellow (bottom) respectively.}
  \label{fig:c2atttime}
\end{figure}

For more particles, we see interference fringes similar to the
repulsive case. We note that the central peak increases and
sharpens with time, indicating increasing contribution from bound
states to the correlations (see Fig.~\ref{fig:3patt} for an example).
\begin{figure}[h]
  \centering \includegraphics[width=8.5cm]{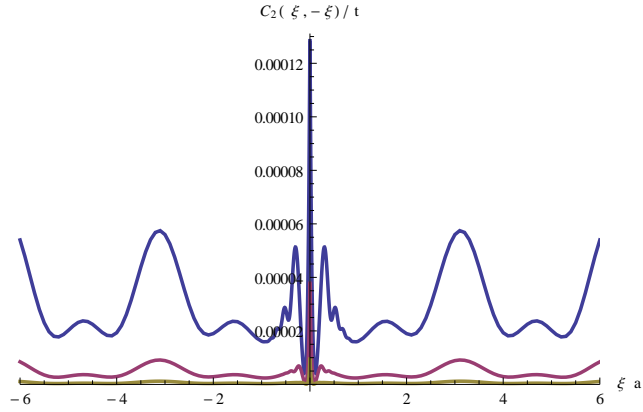}
  \caption{(Color online) $C_2(\xi,-\xi)$ for three particles in the
    attractive case plotted for three different times. At larger
    times, the correlations away from zero fall off. $ta^2=20,40,60$
    for blue (top), magenta (middle) and yellow (bottom) respectively. (from
    Ref.~\onlinecite{deeprl})}
  \label{fig:3patt}
\end{figure}

\hspace*{.1cm}

{\bf  The condensate - attractive and repulsive
  interactions:}  We study the evolution of the Bose gas after a quench
from an initial state where all the bosons are in a single level of a
harmonic trap. For $t<0$, the state is described by
\begin{equation}
  \ket{\Psi_{\text{cond}}} = \int_x \mathcal{S}_x \prod_j \frac{e^{-\frac{x_j^2}{\sigma^2}}}{(\pi\sigma^2)^{\frac14}} b^\dag(x_j)\ket{0}.
\end{equation}
Recall that in order to use the Yudson representation, the initial
state needs to be ordered. We can rewrite the above state as
\bea
  \ket{\Psi_{\text{cond}}} = \int_x   \mathcal{S}_x \theta(x_1>\cdots>x_N)\prod_j
  \frac{e^{-\frac{x_j^2}{\sigma^2}}}{(\pi\sigma^2)^{\frac14}}
  b^\dag(x_j)\ket{0}
\eea
where $\mathcal{S}$ is a symmetrizer. The time evolution can be
carried out via the Yudson representation, and again, we concentrate
on the asymptotics.  For the repulsive model, the stationary phase
contribution is all that appears, and we get
\bea
  \ket{\vec x} = \int_y \prod_{i<j} S^y_{ij}\left(\frac{y_i-y_j-x_i+x_j}{2t}\right)
  \prod_j \frac{1}{\sqrt{2\pi
      it}}e^{-i\frac{(y_j-x_j)^2}{4t}+i\frac{(y_j-x_j)^2}{2t}}
  b^\dag(y_j)\ket{0}.
\eea
At large time $t$, we therefore have
\bea
  \ket{\Psi_{\text{cond}}(t)} = \int_{x,y} \theta(x_1>\cdots>x_N)\phi_2(x) I(y,x,t)  \prod_j b^\dag(y_j)\ket{0}
\eea
$\phi_2(x)$ is symmetric in $x$. $I(y,x,t)$ is symmetric in the $y$
but not in the $x$. Therefore we have to carry out the $x$ integration
over the wedge $x_1>\cdots>x_N$. This is not straightforward to carry
out. If $I(y,x,t)$ was also symmetric in $x$, then we can add the
other wedges to rebuild the full space in $x$. However, due to the
$S$-matrix factors, symmetrizing in $y$ does not automatically
symmetrize in $x$. The exponential factors on the other hand are
automatically symmetric in both variables if one of them is
symmetrized because their functional dependence is of the form
$f(y_j-x_j)$. It is however possible to make the $S$-matrix factors
approximately symmetric in $x$, and we will define what we mean by
approximately shortly. What is important is to obtain a $y_j-x_j$
dependence. As of now, the $S$-matrix that appears in the above
expression is
\begin{equation}
  S^y_{ij}\left(\frac{y_i-y_j-x_i+x_j}{2t}\right) = \frac{\frac{y_i-y_j-x_i+x_j}{2t} -ic\sgn(y_i-y_j)}{\frac{y_i-y_j-x_i+x_j}{2t} -ic}
\end{equation}
First, we can change $\sgn(y_i-y_j)$ to
$\sgn\left(\frac{y_i-y_j}{2t}\right)$ since $t>0$. Next, note that
asymptotically in time, the stationary phase contribution comes from
$\frac{y}{2t}\sim\mathcal{O}(1)$. However, since $x$ has finite
extent, at large enough time, $\frac{x}{2t}\sim 0$. We are therefore
justified in writing
$\sgn\left(\frac{y_i-x_i}{2t}-\frac{y_j-x_j}{2t}\right)$. The only
problem could arise when $y_i\sim y_j$. However, if this occurs, then
the $S$-matrix is approximately $\sgn(y_i-y_j)$ which is antisymmetric
in $ij$. With this prefactor the particles are effectively fermions,
and therefore at $y_i\sim y_j$, the wave-function has an approximate
node. At large time therefore, we do not have to be concerned with the
possibility of particles overlapping, and including the $x_i$ inside
the $\sgn$ function is valid. With this change the $S$-matrix
also becomes a function of $y_j-x_j$ and symmetrizing over $y$ one
automatically symmetrizes over $x$.

In short, we have established that the wave function asymptotically in
time can be made symmetric in $x$. This allows us to rebuild the full
space. We get
\bea
    \ket{\Psi_{\text{cond}}(t)} = \int_{x,y} \sum_P\theta(x_P) \phi_2(x) I^s(y,x,t) \prod_jb^\dag(y_j)\ket{0} = \int_{x,y} \phi_2(x) I^s(y,x,t) \prod_jb^\dag(y_j)\ket{0}
\nn  \eea
where the $s$ superscript indicates that we have established that
$I(y,x,t)$ is also symmetric in $x$. With this in mind, we can do away
with the ordering when we're integrating over the $x$ if we symmetrize
the initial state wave function and the final wave function.  Note
that when we calculate the expectation value of a physical observable,
the symmetry of the wavefunction is automatically enforced, and thus
taken care of automatically.

Recall that when we calculated the noise correlations of the repulsive
gas, in order to get an analytic expression for $N$ particles, we
considered the leading order term, i.e., the HBT term.  We did this by
showing that higher order crossings produced terms higher order in
$e^{-2ca}$ which we claimed was a small number. Now, however, $a=0$,
and although the calculation is essentially the same with our
approximate symmetrization, this simplification does not occur.  The
two and three particle results remain analytically calculable, but for
higher numbers, we have to resort to numerical
integration. Fig.~\ref{fig:c2repcond23} shows the noise correlation
for two and three repulsive bosons starting from a condensate. For
non-interacting particles, we expect a straight line $C_2=
\frac{1}{2}$. When repulsive interactions are turned on, we see the
characteristic fermionic dip develop. The plots for the attractive
Bose gas are shown in Figs.~\ref{fig:c2attcondt} and
\ref{fig:c2attcond3}. As expected from the non-interacting case the
oscillations arising from the interference of particles separated
spatially does not appear. The attractive however does show the
oscillations near the central peak that are also visible in the case
when we start from a lattice-like state. It is interesting to note
that for three particles we do not see any additional structure
develop in the attractive case.

\begin{figure}[htb]
  \centering \includegraphics[width=8.5cm]{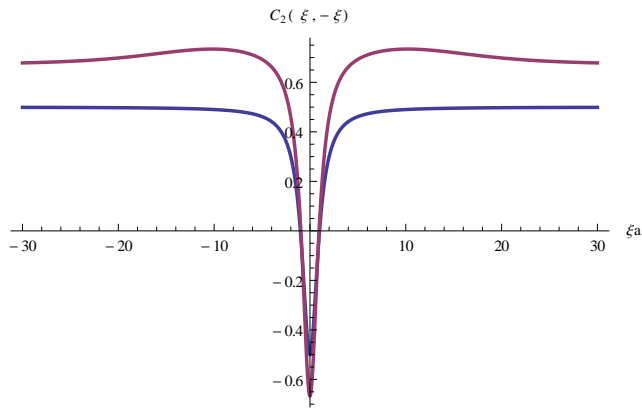}
  \caption{(Color online) $C_2(\xi,-\xi)$ for two (blue, bottom) and three (magenta, top)repulsive bosons
    starting from a condensate. Unlike the attractive case, there is
    no explicit time dependence asymptotically. $ca=3$ (from
    Ref.~\onlinecite{deeprl})}
  \label{fig:c2repcond23}
\end{figure}
\begin{figure}[htb]
  \centering \includegraphics[width=8.5cm]{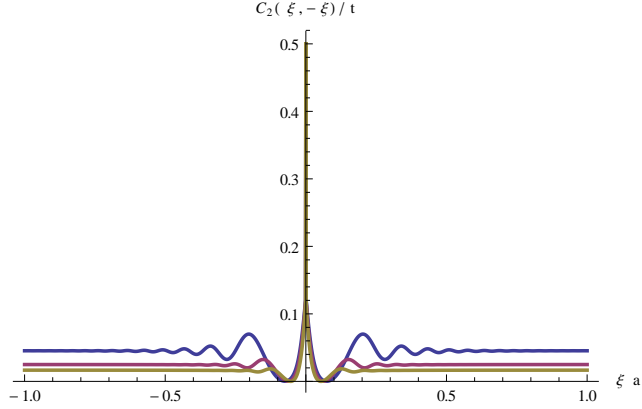}
  \caption{(Color online) Noise correlation for two attractive bosons starting from a
    condensate - as time increases from blue (top) to yellow (bottom), the central peak dominates.}
  \label{fig:c2attcondt}
\end{figure}
\begin{figure}[htb]
  \centering \includegraphics[width=8.5cm]{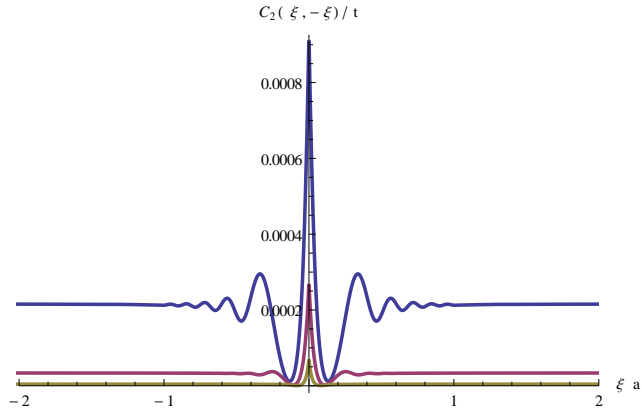}
  \caption{(Color online) $C_2(\xi,-\xi)$ for three attractive bosons starting from a
    condensate. Note that the side peak structure found in
    fig.~\ref{fig:3patt} is missing due to the initial condition. We
    show the evolution at three times. As time increases, the
    oscillations near the central peak die out. Times from top to
    bottom $t c^2=20,40,60$. (from Ref.~\onlinecite{deeprl})}
  \label{fig:c2attcond3}
\end{figure}

\hspace*{.1cm}

{\bf Quenching from a bound state:}
In this brief section our initial state is the ground state of the attractive Lieb-Liniger Hamiltonian (with interaction strength $-c_{0}<0$.
 For two bosons, this take the form~\cite{lieblin},
 \begin{equation}
  \ket{\Psi_{\text{bound}}} = \int_{\vec{x}}e^{-c_{0}\abs{x_{1}-x_{2}} - \frac{x_{1}^{2}}{2\sigma^{2}}-\frac{x_{2}^{2}}{2\sigma^{2}}} \,b^{\dag}(x_{1})b^{\dag}(x_{2})\ket{0},
\end{equation}
and we quench it with a repulsive Hamiltonian.The long time noise correlations are displayed in Fig.~\ref{fig:boundquench}. We see that while the initial state
 correlations are preserved over most of the evolution, in the asymptotic long time limit the characteristic fermionic dip.  We expect similar effects for any number of bosons.
\begin{figure}[htb]
  \centering \includegraphics[width=8cm]{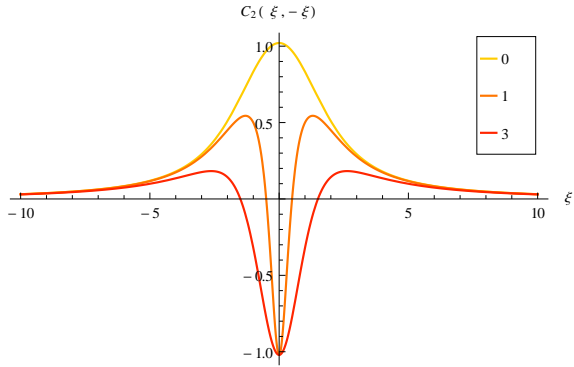}
  \caption{(Color online) Normalized noise correlation function for two particle quenched from a bound state into the repulsive regime.
  The legend indicates the values of $c$ that the state is quenched into. We start with $c_{0}\sigma^{2}=3,\;\sigma=1$,
  $c_{0}$ being the interaction strength of the initial state Hamiltonian.
  Again, we see the fermionic dip, but the rest of the structure is determined by the initial state.}
  \label{fig:boundquench}
\end{figure}

\hspace*{.1cm}

{\bf A scaling argument }  We noticed the as time evolved the system developed strong correlations, enhancing the effective coupling both in
 the attractive and in the repulsive case. This was shown in detailed calculations of the asymptotics carried out by means of saddle point arguments.
Here we proceed to show it via a simpler, though less powerful reasoning. Begin by considering repulsive interactions.
From \eqref{eq:yudrep} we can see by scaling
$\lambda\to\lambda\sqrt{t}$, we get  $Z^y_{ij}(\lambda_i-\lambda_j) \to \sgn(y_i-y_j)+ O\left(\frac{1}{\sqrt{t}}\right)$,
yielding to leading order,
\bea
    \label{eq:claimt} 
      |\Psi_0,t\rangle& \to&
    \int_x \int_{y}\int_{\lambda}\,\theta(\vec{x})\Psi_0(\vec{x}) \prod_j
    \frac{1}{\sqrt{t}}e^{-i\lambda_j^2+i\lambda_j(y_j-x_j)/\sqrt{t}}
    \prod_{i<j} \sgn(y_i-y_j) b^\dag(y_j)\ket{0}  \nn \\
   & &= \int_{x,y,\lambda,k} \,\theta(\vec{x})\Psi_0(\vec{x}) \prod_j
    e^{-i\lambda_j^2t+i\lambda_j(y_j-x_j)}
    e^{-ik_{j}y_{j}}c^{\dag}_{k_{j}}\ket{0} \nn \\
    &&=\int_{x,k} \,\theta(\vec{x})\Psi_0(\vec{x}) \prod_j e^{-ik_j^2t-ik_{j} x_j}c^{\dag}_{k_{j}}\ket{0} \nn \\
    &&= e^{-iH^f_0
      t}\int_{x}\mathcal{A}_x\theta(\vec{x})\Psi_0(\vec{x}) \prod_j
    c^\dag(x_j)\ket{0},
\eea
with $c^{\dagger}(y)$ being fermionic creation operators replacing the
hardcore bosonic operators, $\prod_j
c^{\dagger}(y_j)=\prod_{i<j} \sgn(y_i-y_j) b^\dag(y_j)$. We denote
$H^f_0=\int_x \partial c^{\dagger}(x)\partial c(x)$ the free fermionic
Hamiltonian and $\mathcal{A}_y$ is an anti-symmetrizer acting on the
$y$ variables.  Thus, the repulsive Bose gas, for \emph{any} value of
$c>0$, is governed in the long time by the $c=\infty$ hard core boson
limit (or its fermionic equivalent)~\cite{jukic,girardeau}, and the
system equilibrates (but does not thermalize)  with the antisymmetric wavefunction
$\tilde{\Psi}_0(\vec{y})=\mathcal{A}_y \theta(\vec{y})\Psi_0(\vec{y})$
and the total energy, $E_{\Psi_0}=\langle\Psi_0\rvert H
\lvert\Psi_0\rangle$.  Hence the "fermonic" dip we observed  in the noise correlation function. 

For attractive interaction the argument needs to be refined since contributions of the poles have  to be 
taken into account. They dominate in the long time limit, again corresponding to an effective increase of the attractive 
interaction strength as time evolves.

\section{Conclusions and the dynamic RG hypothesis}
\label{sec:concl}
We have shown that the Yudson contour integral representation for
arbitrary states can indeed be used to understand aspects of the quench
dynamics of the Lieb-Liniger model, and obtain the asymptotic wave functions exactly.
The
representation overcomes some of the major difficulties involved in
using the Bethe-Ansatz to study the dynamics of some integrable
systems by automatically accounting for complicated states in the spectrum.

We see some interesting dynamical effects at long times.  The infinite
time limit of the repulsive model corresponds to particles evolving
with a free fermionic Hamiltonian. It retains, however, memory of the
initial state and
therefore is not a thermal state. The correlation functions approach
that of hard core bosons at long time indicating a dynamical increase
in interaction strength. The attractive model also shows a dynamic
strengthening of the interaction and the long time limit is dominated
by a multiparticle bound state. This of course does not mean that it condenses. In fact the
state diffuses over time, but remains strongly correlated.

We may interpret our results in terms of a ``dynamic RG'' in time.
The asymptotic evolutions of the model both for $c>0$ and for $c<0$
are given by the Hamiltonians $H^*_{\pm}$ with $c\to\pm \infty$
respectively. Accepting the RG logic behind the conjecture one would
expect that there would be basins of attraction around the
Lieb-Liniger Hamiltonian with models whose long time evolution would
bring them close to the "dynamic fixed points" $H^*_{\pm}$. One such
Hamiltonian would have short range potentials replacing the
$\delta$-function interaction that renders the Lieb-Liniger model
integrable. Perhaps, lattice models could be also found in this basin. 
\footnote{ In spite of its similarity to the Lieb-Liniger model, the Bose-Hubbard model is not such a model. 
 The Hamiltonian as defined on the bipartite lattice has a  symmetry that 
leads  attractive and repulsive interaction to evolve identically \cite{deebig}. Such a symmetry is not present in the Lieb-Liniger model.
Adding to the lattice model such terms as the next
nearest hopping or interactions that break this symmetry may lead to Hamiltonians in the same basin of attraction}

We have to emphasize, however, that as these models are not integrable,
we do not expect that they would actually flow to $H^*_{\pm}$. Instead, starting close enough in the ``basin'', they would flow close
to $H^*_{\pm} $ and spend much time in its neighborhood, eventually
evolving into another, thermal state.  We thus conjecture that away
from integrability, a system would approach the corresponding
non-thermal equilibrium, where the dynamics will slow down leading to
a ``prethermal'' state~\cite{berges}. Fig.~\ref{fig:pretherm} shows a
schematic of this.  Such prethermalization behavior has indeed been
observed in lattice models~\cite{kollath}. The system is expected to
eventually find a thermal state.  It is therefore of interest to
characterize different ways of breaking integrability to see when a
system is ``too far'' from integrability to see this effect and in
what regimes a system can be considered as close to
integrability. For a review and background on this subject,
see Ref.~\onlinecite{polkovnikovrmp}.

Further, the flow diagram in Fig.~\ref{fig:pretherm} might have another axis that represents initial states.
Studying the Bose-Hubbard model  \cite{deebig} shows an interesting initial state dependence.
Whereas the sign of the interaction does not affect the quench dynamics,
the asymptotic state depends strongly on the initial state, with a lattice-like
state leading to fermionization, and a condensate-like state retaining bosonic
correlations. The strong dependence on the initial state in the quench dynamics is evident from
eq.~\eqref{eq:quenchoverlaps} and is subject of much debate, in particular as relating to the Eigenstate
Thermalization Hypothesis~\cite{rigol2,srednickitherm,deutsch}.

\begin{figure}[htb]
  \centering
  \includegraphics[width=8.5cm]{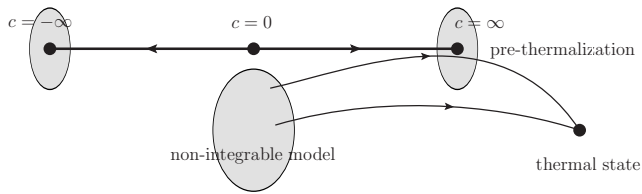}
  \caption{Schematic showing pre-thermalization of states in a
    non-integrable model}
  \label{fig:pretherm}
\end{figure}

We showed here how to compute quantum quenches for the Lieb-Liniger model and provided  predictions for experiments that can be carried out in the context of continuum cold
atom systems.  Application of the approach to other models is underway. This work
 also opens up several new questions: Though  the
representation is provable mathematically, further investigation is
required to understand, physically, how it achieves the tedious sum
over eigenstates, while automatically accounting for the details of
the spectrum. It would also be useful to tie this approach to other means of calculating
overlaps in the Algebraic Bethe Ansatz, i.e., the form-factor approach.
The representation can essentially be thought of as a different way
of writing the identity operator. From that standpoint, it could serve
as a new way of evaluating correlation functions using the Bethe
Ansatz. 

\section{Acknowledgments}
\label{sec:ack}
We are grateful to G.~Goldstein for very
useful discussions. This work was supported by NSF grant DMR 1410583.

\bibliography{LesHouches2}

\end{document}